\documentclass[twocolumn,preprintnumbers,amsmath,amssymb]{revtex4}

\usepackage{graphicx}
\usepackage{dcolumn}
\usepackage{bm}

\begin{document}

\title{Quark-Gluon Bags with Surface Tension}

\author{K. A. Bugaev}
\affiliation{Bogolyubov Institute for Theoretical Physics,
Kiev, Ukraine
}

\date{\today}
\begin{abstract}
The temperature and  chemical potential  dependent  surface tension of bags  is introduced into  the gas of quark-gluon bags model. This resolves a long standing problem  of  a unified description of
the first and second order phase transition with the cross-over. 
Such an approach  is 
necessary to  model the complicated  properties of
quark-gluon plasma and hadronic matter from the first principles of statistical mechanics. 
The suggested model has an exact analytical solution and allows one to rigorously study 
the vicinity of the critical endpoint  of the deconfinement  phase transition. The existence of  higher order phase transitions at the critical endpoint is 
discussed.
In addition, we found that at the curve  of  a zero surface tension coefficient 
 there must exist 
the surface induced  phase tranition of the 2$^{nd}$ or higher order, which separates 
the pure quark gluon plasma (QGP)  from the cross-over states, that are the mixed states of hadrons and QGP bags.  Thus, the present  model predicts that the  critical endpoint of quantum chromodynamics  is the tricritical endpoint.

\vspace*{0.25cm} \hfill  {\it To my mom who taught me how to use  arithmetics}
\end{abstract}

\maketitle


\section{Introduction}

\vspace*{-0.25cm}

Investigation of the strongly interacting matter properties observed in relativistic nuclear collisions has reached the stage when the predictions of the lattice quantum chromodynamics (QCD) can be checked  experimentally on the existing data and 
future mesurements at BNL RHIC, CERN SPS, and GSI FAIR. 
However, a comparison of the theoretical results with the experimental data is not straightforward because during the collision process the matter can have  several  
phase transformations which are  difficult to model. 
 The latter reason stimulated the development of a wide range of phenomenological models of the  strongly interacting matter equation of state  which are used in dynamical simulations. 

One of these models is the famous bag model \cite{BagM} which treats the hadrons as
the bags of  QGP confined inside a hadron with help of   bag pressure.  
The bag model is able to simultaneously describe the hadron mass spectrum, i.e.
the hadron masses and their proper volumes, and the properties of the deconfined phase \cite{bag-qgp}.  This success led to a development of the statistical model 
of QGP, the gas of bags model (GBM) \cite{Goren:81,HR,K}, which 
itself contains  two
well-known models of deconfined and confined phases:
the bag model of   QGP \cite{bag-qgp} and
the hadron gas model \cite{HG}.  
There were hopes \cite{Goren:05}  that an exact analytical solution of the GBM found in 
\cite{Goren:81} could  be helpful in understanding the properties of strongly interacting matter. However, this solution does not allow one to introduce the critical end point 
of  the strongly interacting matter phase diagram.  Also,  a complicated construction 
of the line, along which the phase transition  order  gradually increases, suggested 
in  \cite{Goren:05}, does look too artificial. Therefore, the present GBM formulation 
lacks an important physical input and 
is  interesting only as a toy example which can be solved analytically.

On the other hand,   the models, which can 
correctly reproduce the  expectation \cite{misha,fodorkatz,karsch} 
that the end point of the 1$^{st}$ 
order phase transition  (PT)  line to QGP  should 
be the 2$^{nd}$ order critical point, are indeed  necessary for heavy ion phenomenology. 
In addition, such phenomenological models can provide us with the information 
about the phase structure and equation of state of strongly interacting matter which is located   between the critical endpoint and the region of the 
color superconductivity because such an information is unavailable otherwise. 
Therefore, the present work is devoted to the extension of the GMB. 
We think that  the GMB  can be drastically improved   by the inclusion of such a vitally important element as  the surface tension of the quark-gluon bags.

The dynamical surface tension of the quark-gluon bags was estimated long ago \cite{Jaffe},
but it was never used in statistical description of the equation of state.  Moreover, the estimate of the bag surface tension made in  \cite{Jaffe} is negligible for $u$ and $d$ quarks of and, hence, can be safely neglected in our treatment.
Furthermore, 
although  the influence  of the surface energy 
of the QGP bag properties  in the vacuum was discussed recently \cite{Moretto:0511},
the surface tension of large bags was not included into  statistical description of QGP.
Therefore,   the present paper  is devoted  to the investigation and  analysis of the critical
properties of the model of quark-gluon bags with surface tension (QGBST model hereafter). 

In statistical mechanics there are several exactly solvable cluster  models with 
the 1$^{st}$ order PT which describe the critical point properties  very well.
These models are built on the assumptions that 
the difference of  the bulk  part (or the volume dependent part) of  free energy  
of two phases disappears at  phase equilibrium and that, in addition, 
the difference of the surface part (or the surface tension) of  free energy  vanishes 
at the critical point. 
The most famous of them is  the Fisher droplet model (FDM) \cite{Fisher:64,Elliott:06}
which has been successfully used to analyze the condensation  of a gaseous phase 
(droplets of all sizes)   into a liquid. 
The  FDM has been applied to many different systems,
including nuclear multifragmentation  \cite{Moretto}, nucleation of real fluids \cite{Dillmann},
the compressibility factor of real fluids \cite{Kiang}, bags of the Ising model \cite{Ising:clust}
and percolation  bags \cite{Percolation}.

On the basis of the  statistical multifragmentation  model (SMM) \cite{Bondorf:95} 
commonly used to study  nuclear multifragmentation,  there was recently
formulated a simplified SMM version which was 
solved analytically both for infinite \cite{Bugaev:00,Reuter:01} and 
for finite \cite{Bugaev:04a,Bugaev:05b,Bugaev:05c} volumes of the system. 
In the SMM  the surface tension temperature dependence  differs from that one 
of the FDM,  but it was shown  \cite{Reuter:01} that the value of  Fisher exponent 
$\tau_{SMM} = 1.825 \pm 0.025$, 
which contradicts  to the FDM value $\tau_{FDM}  \approx 2.16$,  
is consistent with ISiS Collaboration data  \cite{ISIS}
and
EOS Collaboration data  \cite{EOS:00}. 
Lately,  our  analytical  results 
\cite{Reuter:01}
were confirmed by the numerical studies \cite{Karnaukhov:tau,Ogul:tau}.

Such an   experimentally obtained range of the $\tau$ index is of a principal importance because  it gives a very strong evidence that the SMM, and, thus, 
the nuclear matter,  has a tricritical endpoint rather than a critical  endpoint 
\cite{Bugaev:00,Reuter:01}.

This  success of  the SMM initiated  
the studies of the surface partitions of large clusters  
within the Hills and Dales Model
 \cite{Bugaev:04b,Bugaev:05a} and led to a discovery of the origin  of
the temperature independent surface entropy similar to the FDM.  
As a consequence, the surface tension coefficient of large 
clusters consisting of the discrete constituents should linearly depend 
on the temperature of the system \cite{Bugaev:04b} and  must vanish at the critical endpoint.
However, the present formulation of the Hills and Dales Model
\cite{Bugaev:04b,Bugaev:05a}, which successfully 
estimates the upper and lower bounds of the surface deformations of the discrete 
physical clusters, does not look  suitable for  quark-gluon bags.
Therefore, in this work we  assume a certain  dependence of the surface 
tension coefficient on temperature and baryonic chemical potential,  and  concentrate  on the impact  of 
surface tension of the quark-gluon bags on  the properties of 
the deconfinement  phase diagram and 
the QCD critical endpoint. 
A discussion of the origin of the surface tension
is a subject of our future work.

Here we will show that the existence of a cross-over at  low values of
the  baryonic chemical potential along with  the 1$^{st}$ order deconfinement PT
at high baryonic chemical potentials  leads to the existence of an additional PT 
of the 2$^{nd}$ or higher order  along the curve where the surface tension coefficient  vanishes. Thus, it turns out that the QGBST model predicts 
the existence of the tricritical rather than critical endpoint.

The paper  is organized as follows.
Sect. II contains the formulation of the basic ingredients of the GBM. 
In Sect. III we  formulate
the QGBST model  and  analyze  all possible singularities of its  isobaric partition for vanishing
baryonic densities. This analysis is generalized to non-zero  baryonic densities in Sect. IV.
 Sect. V  is devoted to the analysis of the surface tension induced  PT which exists above the deconfinement PT. 
The conclusions and research perspectives are summarized in Sect. V.

\vspace*{-0.25cm}

\section{Basic Ingredients of the  GBM }

\vspace*{-0.25cm}

To remind the basic ingredients of the GBM let us consider 
the Van der Waals gas consisting of $n$ hadronic  species,
which are called bags in what follows, at zero baryonic chemical potential. Its grand canonical partition (GCP)  is 
given by \cite{Goren:81}
\begin{eqnarray}
 Z  (V,T) & =  & \hspace*{-0.2cm}
 \sum_{\{N_k\}} \biggl[
\prod_{k=1}^{n}\frac{\left[ \left( V -v_1N_1-...-v_nN_n\right)  \phi_k(T) \right]^{N_k}}{N_k!} \biggr]  \nonumber \\
%
~& \times &  \theta\left(V-v_1N_1-...-v_nN_n\right)~, 
\label{GCPzeromu}
\end{eqnarray}
where the  function $\phi_k(T) \equiv g_k ~ \phi(T,m_k) $ 
$$\phi_k(T)   \equiv  \frac{g_k}{2\pi^2} \int\limits_0^{\infty}\hspace*{-0.1cm}p^2dp~
e^{\textstyle - \frac{(p^2~+~m_k^2)^{1/2}}{T} }
=  g_k \frac{m_k^2T}{2\pi^2}~{ K}_2\left( \frac{m_k}{T} \right)
$$ 
is the particle  density
of  bags of mass $m_k$ and eigen volume $v_k$  and degeneracy $g_k$. 
Using the standard technique of 
the Laplace transformation \cite{Goren:81,Bugaev:00}  with respect to volume,  
one obtains the isobaric partition:
\begin{align}\label{Zs}
& \hat{Z}(s,T) \equiv \int\limits_0^{\infty}dV\exp(-sV)~Z(V,T) =\frac{1}{ [ s - F(s, T) ] }  \\
& {\rm with} \quad F(s, T) \equiv \sum_{j=1}^n \exp\left(-v_j s\right)~g_j\phi(T,m_j) ~.
\label{FsTi}
\end{align}
From the definition of pressure in the grand canonical ensemble
it follows that, in the thermodynamic limit,
the GCP of the system  behaves as $Z(V,T)\simeq \exp\left[pV/T \right]$.
An exponentially increasing $Z(V,T)$ generates the rightmost singularity $s^*=p/T$ 
of the function $\hat{Z}(s,T)$ in variable $s$. 
This is because the integral over $V$ in Eq.~(\ref{Zs}) 
diverges at its upper limit for $s < p/T$. 
Therefore, the rightmost singularity 
$s^*$ of $\hat{Z}(s,T)$ gives us the system  pressure:
\begin{align}\label{p-s}
p(T)~=~T~\lim_{V\rightarrow\infty}\frac{\ln Z(V,T)}{V}~=~T~s^*(T)~.
\end{align}
The  singularity 
$s^*$ of $\hat{Z}(s,T)$ (\ref{Zs}) can be calculated from the transcendental
equation \cite{Goren:81,Bugaev:00}:
\begin{align}\label{s*vdw}
s^*(T)~=~ F(s^*,T)~.
 \end{align}
As long as the number of bags, $n$, is finite, the only possible singularities  
of $\hat{Z}(s,T)$ (\ref{Zs}) are simple   poles. 
For example, for the ideal gas ($n = 1; v_1=0$ in  Eq.~(\ref{s*vdw}))
$s^*=g_1\phi(T,m_1)$ and thus from  Eq.~(\ref{p-s}) one gets
$p=Tg_1\phi(T,m_1)$ which corresponds to the  grand 
canonical ensemble ideal gas equation of state
for the particles of  mass $m_1$ and degeneracy $g_1$.

However, in the case of an infinite number of sorts of  bags  an essential  singularity of
$\hat{Z}(s,T)$ may appear.  
This  property is used  in  the GBM: 
to the finite sum  over different bag states in (\ref{Zs})
 the integral $\int_{M_0}^{\infty}dm\, dv ...\rho(m,v)$ is added 
 with  the bag mass-volume spectrum, $\rho(m,v)$,  which defines  
the number of bag states in the mass-volume  region  $[m,v;m+dm,v+dv]$.
In this case  the function $F(s,T)$ in Eqs. (\ref{Zs}) and (\ref{s*vdw}) 
should be replaced by
\begin{align}
F(s,T)&\equiv F_H(s,T)+F_Q(s,T) = \sum_{j=1}^n g_j e^{-v_js} \phi(T,m_j) \nonumber \\
&+\int\limits_{V_o}^{\infty}dv\hspace*{-0.3cm}\int\limits_{M_o+Bv}^{\infty}
 \hspace*{-0.3cm}dm~\rho(m,v)\exp(-sv)\phi(T,m)~.
 \label{FsHQ}
\end{align}
The first term of Eq.~(\ref{FsHQ}), $F_H$, represents the contribution of a finite number of low-lying
hadron states. This function has no $s$-singularities at
any temperature $T$ and can generate a simple pole of the isobaric partition, whereas  the mass-volume spectrum of the bags $F_Q(s,T)$ can be chosen to 
generate an essential  singularity $s_Q (T) \equiv p_Q(T)/T$ which defines  the QGP  pressure $p_Q(T)$  at zero baryonic densities \cite{Goren:81,Goren:82}. 

The mass-volume spectrum is the generalization of the exponential mass spectrum 
introduced by Hagedorn \cite{Hagedorn:65,Hagedorn:68}. 
The usage of the grand canonical  description for  the exponential mass spectrum  
was recently strongly criticized \cite{Moretto:06,BugaevMoretto:05a,Elliott:05}  because  of the thermostatic properties of this spectrum.
Fortunately, the Van der Waals repulsion  compensates  the growing part of the mass-volume spectrum and, hence, the criticism of Refs. 
\cite{Moretto:06,BugaevMoretto:05a,Elliott:05} is irrelevant to the present model. 

There are several possibilities to parameterize the mass-volume spectrum $\rho(m,v)$.
Thus, in the simplest case one can assume that for heavy resonances their mass and 
eigen volume are proportional, i.e. the spectrum $\rho(m,v)$ contains the function
$\delta(m - v\, Const)$. An alternative choice was suggested in \cite{Goren:82},
but in either case the resulting expression for  the continuum spectrum of the 
GBM $F_Q(s,T)$ can be cast as
\begin{align}\label{FQs}
&F_Q(s,T)~ = ~u(T)
 ~  \int\limits_{V_0}^{\infty}dv~\frac{ \exp\left[-v\left(s -s_Q(T)\right)\right] }{v^{\tau}}~,
 \end{align}
where
$u(T)$, $\tau > 0$ are the model  parameters. The QGP  pressure  $p (T) = T s_Q(T)$
can be parameterized  in many ways. For instance, the MIT bag model equation of state
\cite{BagM} corresponds \cite{Goren:82} to  $s_Q(T) \equiv  \frac{1}{3}~\sigma_Q~T^3~-~\frac{B}{T} $ and
$u(T) =C\pi^{-1}\sigma_Q^{\delta+1/2}~T^{4+4\delta}~\left(\sigma_QT^4+B\right)^{3/2}$.
Here $B$ denotes the bag constant, $\sigma_Q~=~\frac{\pi^2}{30}~\frac{95}{2}$
is the Stefan-Boltzmann constant counting gluons (spin, color) and (anti-)quarks
(spin, color and $u$, $d$, $s$-flavor) degrees of freedom; and the constants $C$,  
$ \delta <  0$, $V_o \approx 1$ fm$^3$ and $M_o \approx 2$ GeV are the parameters
of the mass-volume spectrum.

\vspace*{-0.25cm}

\section{The Role of Surface Tension}

\vspace*{-0.25cm}

At the moment the particular choice of function $F_Q(s,T)$ (\ref{FQs}) is not important. 
The key point for our study is that it should have the form of Eq.  (\ref{FQs}) which 
has a singularity  at  $s=s_Q$ 
because for $s<s_Q$ the integral over $dv$  diverges at its upper limit. 
Note that 
the exponential in (\ref{FQs}) is nothing else, but a difference of the bulk 
free energy of a bag of volume $v$, i.e. $ -T s v$,  which is  under external pressure 
$T s$,   and  the bulk  free energy of 
the same bag filled with QGP, i.e.  $ -T s_Q v$. 
At phase equilibrium this difference of the bulk free energies  vanishes. 
Despite  all  positive features,  Eq. (\ref{FQs}) lacks the surface part of  free energy of bags, which will be called  a surface energy hereafter. 
In addition to the difference of the bulk free energies
the realistic statistical models which demonstrated  their validity, 
the FDM \cite{Fisher:64} and SMM \cite{Bondorf:95}, 
have the contribution of the surface  energy which plays an important role in  defining the  phase diagram structure \cite{Bugaev:00,Bugaev:05c}. 
Therefore, we modify Eq. (\ref{FQs}) by introducing the surface   energy of the bags in a general fashion \cite{Reuter:01}:
\begin{align}\label{FQsnew}
&F_Q = u(T)
   \int\limits_{V_0}^{\infty}dv~\frac{ \exp\left[\left(s_Q(T)-s\right)v - \sigma(T)\, v^{\varkappa}\right] }{v^{\tau}}\,,
 \end{align}
where the ratio of the  temperature dependent surface tension coefficient  to $T$
(the reduced surface tension coefficient hereafter) 
which has the form $\sigma(T) = 
\frac{\sigma_o}{T} \cdot
\left[ \frac{ T_{cep}   - T }{T_{cep}} \right]^{2k + 1} $  ($k =0, 1, 2,...$).  
Here $\sigma_o > 0$ can be a smooth function of the temperature, but for simplicity we fix it to be a constant.  
For $k = 0$ the two terms in the surface (free) energy of a $v$-volume bag  have 
a simple interpretation \cite{Fisher:64}: thus, the surface energy of such a bag is
$\sigma_0 v^{\varkappa}$, whereas the free energy, which comes from  the surface entropy $\sigma_o T_{cep}^{-1} v^{\varkappa} $,  is  
$- T \sigma_o T_{cep}^{-1} v^{\varkappa}$.  
Note that   the surface entropy of a  $v$-volume  bag
counts its degeneracy factor or the number of ways to make 
such a bag with all possible surfaces. This   interpretation 
can be extended to 
$k >  0$  on the basis of  the Hills and Dales Model 
\cite{Bugaev:04b,Bugaev:05a}.

In choosing such a simple surface energy parameterization we 
follow the original Fisher idea \cite{Fisher:64}  which allows one to account for 
the surface energy by considering 
some mean bag of volume $v$ and surface $v^{\varkappa}$. The 
consideration of 
the general mass-volume-surface bag spectrum we leave  for the future investigation. 
The power  $\varkappa < 1$ which describes the bag's effective  surface is a constant which,  in principle, can differ from the typical FDM and SMM value 
$\frac{2}{3}$.
This is so because  near  the deconfinement PT region  QGP  has  low density and, hence, 
like in the low density  nuclear matter \cite{Ravenhall},  
the non-sperical bags (spaghetti-like or lasagna-like \cite{Ravenhall})  can be  favorable.
A similar idea of  ``polymerization" of gluonic quasiparticles was introduced recently 
\cite{Shuryak:05a}.

The second  essential  difference with the FDM and SMM surface tension 
parameterization is that we do not require the  vanishing of $\sigma(T)$ above the CEP. 
As will be shown later,  this is the most important assumption which, in contrast to the GBM,  allows one to naturally describe the cross-over  from hadron gas to QGP.  
Note that  negative value of the reduced surface tension coefficient $\sigma(T)$ above the CEP 
does not mean anything wrong. As we discussed above, 
the  surface  tension coefficient consists of energy and entropy parts which  have   opposite signs \cite{Fisher:64,Bugaev:04b,Bugaev:05a}. 
Therefore, $\sigma(T) < 0 $ does not mean that the surface energy changes the sign, but it
rather  means that the surface entropy, i.e. the logarithm of the degeneracy of bags of a fixed volume, simply  exceeds their  surface energy.  In other words, 
the number of  non-spherical bags of a fixed volume becomes so large that the Boltzmann exponent, which accounts for the energy "costs" of these bags,  cannot
suppress them anymore.

Finally, the third essential difference with the FDM and SMM is that we assume 
that the surface tension in the QGBST model happens at some line in $\mu_B-T$ plane, 
i.e. $T_{cep} = T_{cep} (\mu_B)$. However,  in the subsequent  sections we will 
consider $T_{cep} = Const $ for simplicity, and in Sect. V  we  will  discuss the necessary modifications of the model  with $T_{cep} = T_{cep} (\mu_B)$.

The surface energy should, in principle, be introduced into a  discrete part of 
the mass-volume spectrum $F_H$, but a successful  fitting of the particle yield ratios \cite{HG} with 
the experimentally determined hadronic spectrum $F_H$ does not indicate 
such a necessity.

According to the general theorem \cite{Goren:81} the analysis of PT  existence of the GCP  is now reduced to the analysis of the rightmost singularity of 
the isobaric partition (\ref{Zs}).  Depending on the sign of the reduced surface tension coefficient, there are three possibilities. 

\noindent
({\bf I}) The first possibility corresponds to $\sigma(T) > 0$. Its treatment is  very similar 
to the GBM choice (\ref{FQs})  with $\tau > 2$ \cite{Goren:81}. In this case at low 
temperatures the QGP  pressure $T s_Q(T)$ is negative and, therefore, the   rightmost singularity is a simple pole of the isobaric partition  
$s^* = s_H (T) = F(s_H(T), T) > s_Q(T)$, which is mainly defined by a discrete part of  the mass-volume spectrum $F_H(s,T)$. 
The last inequality provides the convergence of the volume integral in (\ref{FQsnew})
(see Fig. 1). On the other hand at very high $T$ the QGP pressure dominates and, hence, the rightmost singularity is the essential  singularity  of the isobaric partition  $ s^* = s_Q(T)$.  
The phase transition occurs, when the singularities coincide:
\begin{align}\label{PTI}
&  s_H (T_c) \equiv  \frac{p_H (T_c)}{T_c} =  s_Q (T_c) \equiv  \frac{p_Q (T_c)}{T_c}\,,
 \end{align}
which is nothing else, but the Gibbs criterion. 
The graphical solution of Eq. (\ref{s*vdw})  for  all  these possibilities is shown in Fig. 1.  
Like in the GBM \cite{Goren:81,Goren:05}, the  necessary condition for the PT existence is  the finiteness of $F_Q(s_Q(T), T) $ at $s = s_Q(T)$.
It can be shown that the sufficient conditions are   the following 
inequalities: $F_Q(s_Q(T), T) > s_Q(T) $ for low temperatures and 
  $F(s_Q(T), T) < s_Q(T) $  for  $T \rightarrow \infty$.  These 
conditions provide that at low $T$ the rightmost singularity  of the isobaric partition is a simple pole,
whereas for hight $T$ the essential singularity 
 $s_Q(T) $ becomes  its  rightmost one   (see Fig. 1 and a detailed 
analysis of case $\mu_B \neq 0$).

%
%
\begin{figure}[ht]
\includegraphics[width=6.3cm,height=6.3cm]{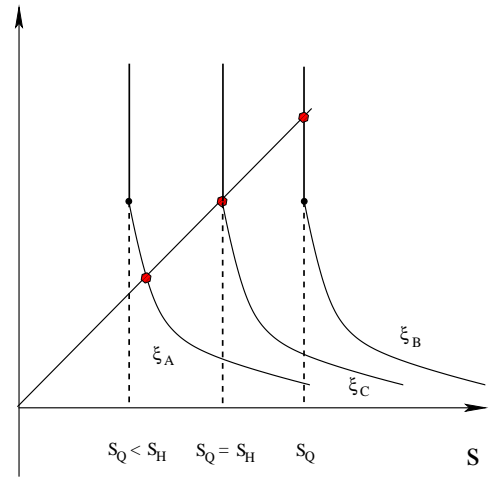}
\vspace*{-0.3cm}
\caption{
Graphical solution of Eq. (\ref{s*vdw}) which corresponds to a PT.
The solution of Eq. (\ref{s*vdw}) is shown by a filled hexagon.
The function $F(s, \xi)$ is shown by a solid curve for a few 
values of  the parameter $\xi$.  The function  $F(s, \xi)$ diverges 
for $s < s_Q( \xi)$ (shown by dashed lines), but is finite at $s = s_Q( \xi)$ . 
At low values of  the parameter $\xi = \xi_A$, which can be either $T$ or $\mu_B$, 
the simple pole $s_H$ is the rightmost singularity and it corresponds to hadronic phase. 
For  $\xi = \xi_B \gg \xi_A$ the  rightmost singularity is an essential singularity $s = s_Q( \xi_B)$, 
which describes QGP. 
At intermediate  value $\xi = \xi_C$ both singularities coincide $s_H( \xi_C) = s_Q( \xi_C)$ and 
this condition is a Gibbs criterion. 
}
  \label{fig1}
\end{figure}

\vspace*{-0.0cm}

The PT order can be found from the $T$-derivatives of  $s_H (T)$. 
Thus,  differentiating  (\ref{s*vdw}) one finds 
\vspace*{-0.2cm}
\begin{align}\label{sHprime1}
s_H^{\prime}~=~
\frac{G~+~u\,{\cal K}_{\tau-1}(\Delta, - \sigma) \cdot  s_Q^{\prime}}{1~+~u\,{\cal K}_{\tau-1}(\Delta, - \sigma)} \,,
\end{align}

\vspace*{-0.2cm}

\noindent
where the functions $G $ and ${\cal K}_{\tau -a} (\Delta, - \sigma) $ are defined as
\begin{align}\label{G1}
&\hspace*{-0.25cm}G \equiv F_H^{\prime}+ \frac{u^{\prime}}{u} F_Q 
+  {\textstyle\frac{ (T_{cep} - 2k T) \sigma(T)}{(T_{cep} - T)\, T } }\,u \,
{\cal K}_{\tau- \varkappa}  (\Delta, - \sigma)\,, \\
\label{KQ}
&\hspace*{-0.25cm}{\cal K}_{\tau -a} (\Delta, - \sigma) \equiv  \hspace*{-0.0cm}
\int\limits_{V_o}^{\infty}\hspace*{-0.05cm}dv~\frac{\exp\left[-\Delta v - \sigma(T) 
v^{\varkappa} 
\right] }{v^{\tau-a}} \,,
\end{align}
where $\Delta \equiv s_H - s_Q$. 

Now it is easy to see that the transition is of the 1$^{st}$ order,
i.e. $s_Q^{\prime}(T_c)>s_H^{\prime}(T_c)$, provided  $ \sigma(T) > 0$ for any $\tau$.
The 2$^{nd}$ or higher order phase transition takes place
provided $s_Q^{\prime}(T_c)=s_H^{\prime}(T_c)$ at $T=T_c$.
The latter condition is satisfied  when $ {\cal K}_{\tau-1}$ diverges to infinity
at $T\rightarrow (T_c-0)$, i.e. for $T$ approaching $T_c$ from below.
Like for the GBM choice (\ref{FQs}), 
such a situation can exist for   $ \sigma(T_c) = 0$ and $\frac{3}{2} < \tau \le  2 $. 
Studying the higher $T$-derivatives  of $s_H(T)$ at $T_c$, one can show 
that  for  $ \sigma(T) \equiv  0$  and  for $(n+1)/n \le \tau < n/(n-1)$ ($n=3,4,5,...$) there is a $n^{th}$ order phase  transition
\begin{align}\label{nth}
& s_H(T_c)~ =~ s_Q(T_c)~,~~
s_H^{\prime}(T_c)~ =~ s_Q^{\prime}(T_c)~,~...~ \nonumber \\
& s_H^{(n-1)}(T_c)~=~ s_Q^{(n-1)}(T_c)~,~~
s_H^{(n)}(T_c)~\ne~ s_Q^{(n)}(T_c)~,
\end{align}
 with $ s_H^{(n)}(T_c)=\infty$
for $(n+1)/n < \tau < n/(n-1)$
and  with a finite value of $s_H^{(n)}(T_c)$ for $\tau =(n+1)/n$.\\

%
%
\begin{figure}[ht]
\includegraphics[width=6.3cm,height=6.3cm]{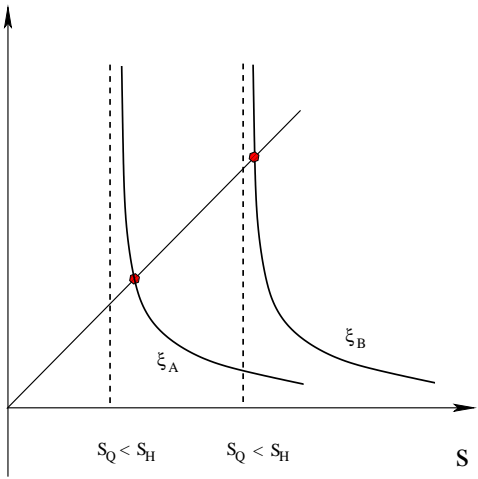}
\vspace*{-0.3cm}
\caption{
Graphical solution of Eq. (\ref{s*vdw}) which corresponds to a cross-over.
The notations are the same as in Fig. 1. 
Now the function  $F(s, \xi)$ diverges 
at $s = s_Q( \xi)$ (shown by dashed lines). 
In this case the simple pole $s_H$ is the rightmost singularity for any value of $\xi $. 
}
\label{fig2}
\end{figure}

\vspace*{-0.0cm}

\noindent
({\bf II}) The second possibility, $\sigma(T) \equiv 0$, described in the preceding paragraph,  does not give anything new 
compared to the GBM \cite{Goren:81,Goren:05}. 
If the  PT exists, then
the graphical picture of  singularities is basically similar to Fig. 1. The only difference
is that, depending on the PT order,   the derivatives of  $F(s,T) $ function with respect to $s$  should  diverge at $s = s_Q(T_c)$.\\

\noindent
({\bf III})  A principally new possibility exists for $T > T_{cep}$, where $\sigma(T) < 0$.
In this case there exists a  cross-over.  Its existence can be shown as follows. 
Let us solve the equation for singularities (\ref{s*vdw}) graphically (see Fig. 2). 
For  $\sigma(T) < 0$ the function $F_Q(s,T)$ diverges at $s = s_Q(T)$. 
On the other hand, the partial derivatives $\frac{\partial F_H(s,T)}{\partial s} < 0$ and 
$\frac{\partial F_Q(s,T)}{\partial s} < 0$ are always negative. Therefore, the  function 
$F(s,T) \equiv  F_H(s,T) +  F_Q(s,T)$ is a monotonically decreasing function of 
$s$, which vanishes at $s \rightarrow \infty$. Since the left hand side of Eq.  (\ref{s*vdw})  is a  monotonically increasing function of $s$, then there can exist a single intersection 
$s^*$ of $s$ and $F(s,T)$ functions. Moreover, for finite  $s_Q(T)$ values this 
intersection can occur  on 
the right hand side of the point $s = s_Q(T)$, i.e.  $s^* > s_Q(T)$ (see Fig. 2). 
Thus, in this case the essential singularity $s = s_Q(T)$ can become the rightmost one
for infinite temperature only.  In other words, the pressure of the pure QGP can be reached 
at infinite $T$, whereas for finite $T$ the hadronic mass spectrum gives a non-zero  contribution into all thermodynamic functions. 

Note also that all these nice properties would vanish, if  the reduced surface tension coefficient is  zero or positive above $T_{cep}$. This is  one of the crucial points of the present model which puts forward certain doubts about the vanishing of 
the reduced  surface tension coefficient  in the FDM 
\cite{Fisher:64} and SMM \cite{Bondorf:95}. These doubts are also supported by the first principle results obtained by the Hills and Dales Model  \cite{Bugaev:04b,Bugaev:05a}, because the surface entropy simply  counts the degeneracy of a cluster of a fixed volume and  it  does not physically affect  the surface energy of this cluster. 

\vspace*{-0.25cm}

\section{Generalization to Non-Zero Baryonic Densities}

\vspace*{-0.25cm}

The possibilities  ({\bf I})-({\bf III}) discussed in the preceding section 
remain unchanged for non-zero baryonic numbers. The latter should be 
included into consideration  to make our model more realistic. To keep 
the presentation simple, we do not account for  strangeness. 
The inclusion of the baryonic charge of the quark-gluon bags 
does not change the two types of singularities of the isobaric partition (\ref{Zs})
and the corresponding equation for them (\ref{s*vdw}), but it 
leads to 
the following modifications of the $F_H$ and $F_Q$ functions:
\begin{align}\label{FHTmu}
F_H&(s,T,\mu_B)= \sum_{j=1}^n g_j     e^{\frac{b_j \mu_B}{T} -v_js} \phi(T,m_j)\,,
\\
F_Q &(s,T,\mu_B) = {\textstyle u(T, {\mu_B})} \nonumber \\
   &\times \int\limits_{V_0}^{\infty}dv~ \frac{ \exp\left[\left(s_Q(T,\mu_B)-s\right)v - \sigma(T) 
v^{\varkappa}\right] }{v^{\tau}}\,. 
\label{FQTmu}   
\end{align}
Here the baryonic chemical potential is denoted as $\mu_B$,  the baryonic charge of 
the $j$-th hadron in the discrete part of the spectrum is $b_j$. The continuous part 
of the spectrum, $F_Q$ can be obtained from  some spectrum $\rho(m,v, b)$
in the spirit of Ref. \cite{Goren:82}, but this will lead us away from the main 
subject. 

The QGP pressure $p_Q = T s_Q(T,\mu_B)$ can be also chosen in several ways. 
Here we  use the bag model pressure 
\begin{align}\label{sQB}
&\hspace*{-0.2cm}p_Q = \frac{\pi^2}{90}T^4 \left[
 \frac{95}{2} +
\frac{10}{\pi^2} \left(\frac{\mu_B}{T}\right)^2 + \frac{5}{9\pi^4}
\left(\frac{\mu_B}{T}\right)^4 \right]
- B \,, 
\end{align}
but the more complicated model pressures, even with the
PT of other kind like the transition between the color superconducting QGP 
and the usual QGP, can be, in principle,  used.

The sufficient conditions for a PT  existence are  
\begin{align}\label{SufCondI}
&\hspace*{-0.cm}
F (s_Q(T,\mu_B=0),  T,\mu_B=0)  >  s_Q(T,\mu_B=0) \,, \\
&\hspace*{-0.0cm}
F (s_Q(T,\mu_B ),  T,\mu_B)  < s_Q(T,\mu_B)\,, ~{\rm for~all} ~ \mu > \mu_A.
\label{SufCondII}
\end{align}
The  condition (\ref{SufCondI})  provides that the simple pole singularity 
$s^* = s_H(T,\mu_B=0)$ is the rightmost 
one at vanishing $\mu_B=0$ and given $T$, whereas  the condition (\ref{SufCondII}) 
ensures that $s^* = s_Q(T,\mu_B)$ is the rightmost singularity of the isobaric partition for 
all values of the baryonic chemical potential above some positive  constant $\mu_A$. 
This can be seen in Fig. 1 for  $\mu_B$ being a variable.  
Since  $F (s, T,\mu_B)$, where it exists,  is a continuous function of its  parameters,
one concludes that, if the conditions (\ref{SufCondI}) and (\ref{SufCondII}), are fulfilled,
then at some chemical potential $\mu_B^c (T)$ the both singularities should  
be equal. Thus, one arrives at the Gibbs criterion (\ref{PTI}), but for two variables
\begin{align}\label{PTII}
&  s_H (T, \mu_B^c(T))  =  s_Q (T, \mu_B^c(T)) \,.
 \end{align}

\noindent 
It is easy to see that the  inequalities  (\ref{SufCondI}) and (\ref{SufCondII}) are  the  sufficient conditions  of a PT existence
for  more complicated functional dependencies of $F_H (s,  T,\mu_B)$ and 
$F_Q (s,  T,\mu_B)$ than the ones used here. 

\vspace*{-0.0cm}

%
%
\begin{figure}[ht]
\includegraphics[width=8.4cm,height=6.3cm]{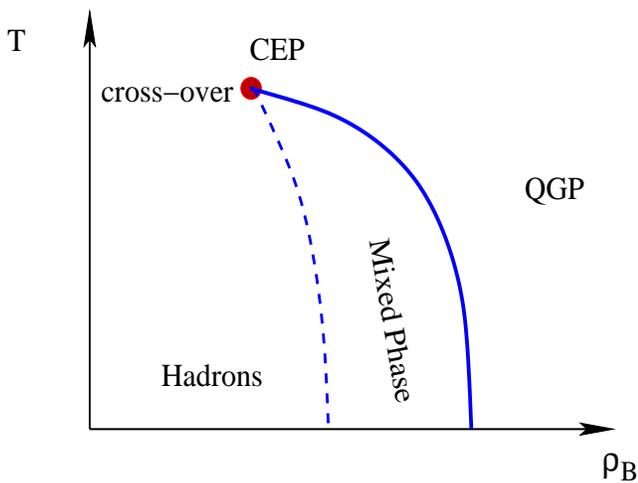}
\vspace*{-0.3cm}
\caption{
A schematic picture of the deconfinement phase transition diagram 
in the plane of baryonic density $\rho_B$  and  $T$ for 
the 2$^{nd}$ order PT at the critical endpoint (CEP), i.e. for $\frac{3}{2} < \tau \le 2$. 
For the 3$^{rd}$ (or higher) order PT  the  boundary of the mixed 
and hadronic phases (dashed curve) should have the same slope as 
the boundary of the mixed phase and QGP (solid curve) at the CEP. 
}
  \label{fig3}
\end{figure}

\vspace*{0.0cm}

For our choice (\ref{FHTmu}), (\ref{FQTmu}) and (\ref{sQB})  of 
$F_H (s,  T,\mu_B)$ and $F_Q (s,  T,\mu_B)$ functions the  PT exists at 
$T < T_{cep}$,  because the sufficient conditions (\ref{SufCondI}) and (\ref{SufCondII}) 
can be  easily fulfilled   by a proper choice of the bag constant $B$ and the function 
${\textstyle u(T, \mu_B)} > 0$ for the interval $T \le T_{up}$ with the constant $T_{up} > T_{cep}$. 
Clearly, this is the  1$^{st}$ order PT, since the surface
tension is finite and it provides the convergence of the  integrals (\ref{G1}) and  (\ref{KQ})
in the expression (\ref{sHprime1}), where  the  usual $T$-derivatives should be now 
understood as the partial ones for $\mu_B = const$.  

Assuming that the conditions (\ref{SufCondI}) and (\ref{SufCondII}) 
 are fulfilled by the correct choice of the  model parameters $B$ and  
 ${\textstyle u(T, \mu_B)} > 0$,   one  can see now that at $T = T_{cep}$ there exists 
a PT as well, but its order is defined by the value of $\tau$. As was discussed in the 
preceding section for  $\frac{3}{2} < \tau \le 2$ there  exists the 2$^{nd}$ order PT. 
For  $1 < \tau \le \frac{3}{2}$  there  exist  the  PT of higher order, defined 
by the conditions formulated in Eq.  (\ref{nth}). 
This is a new possibility, which, to our best knowledge,  does not contradict  to any 
general physical principle (see Fig. 3). 

The case $ \tau > 2$ can be ruled out because  there must  exist
the first order PT for  $T \ge T_{cep}$, whereas for  $T < T_{cep}$ there exists 
the cross-over. Thus,  the critical endpoint in $T-\mu_B$ plane  will correspond to 
the critical  interval  in the   temperature-baryonic density plane.  
Since such a structure 
of the  phase diagram in the variables temperature-density  has, to our knowledge,
never  been observed, we conclude that  the case  $ \tau > 2$ is unrealistic (see Fig. 4). 
Note that a similar phase diagram exists in the FDM with the only  difference  that 
the boundary of the mixed and liquid phases (the latter in the QGBST model corresponds 
to QGP) is moved to infinite particle density.

\vspace*{0.4cm}

%
%
\begin{figure}[ht]
\includegraphics[width=8.4cm,height=6.3cm]{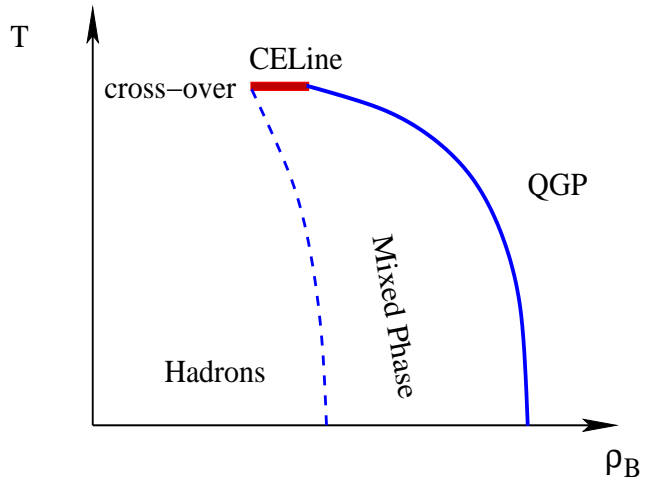}
\vspace*{-0.3cm}
\caption{
A schematic picture of the deconfinement phase transition diagram 
in the plane of baryonic density $\rho_B$  and  $T$ for $ \tau >  2$.
The critical endpoint in the $\mu_B-T$ plane generates 
the critical end line (CELine) in the $\rho_B-T$ plane shown by the thick horizontal line. 
This occurs because of the discontinuity of the partial derivatives of $s_H$ and $s_Q$ 
functions with respect to $\mu_B$ and $T$. 
}
  \label{fig4}
\end{figure}

\vspace*{-0.25cm}

\section{Surface Tension Induced Phase Transition}

\vspace*{-0.25cm}

Using our results for the  case ({\bf III}) of the preceding section, we conclude that 
above $T_{cep}$ there is a cross-over, i.e. the QGP and hadrons coexist together 
up to the infinite values of $T$ and/or $\mu_B$. Now, however, it is necessary to  answer
the question: How can the two different sets of singularities  that exist on two sides of 
the line $T =  T_{cep}$ provide the continuity of the solution of Eq. (\ref{s*vdw})?

It is easy to  answer  this question for $\mu_B < \mu_B^c(T_{cep})$  because 
in this case all partial $T$ derivatives of  $s_H(T, \mu_B) $, which is the rightmost singularity, 
exist and are finite at any point of the line $T =  T_{cep}$.  This can be seen from the
fact that  for the considered  region of parameters  $s_H(T, \mu_B) $ is the rightmost singularity and, consequently, $s_H(T, \mu_B) > s_Q (T, \mu_B)$. The latter inequality 
provides the existence and finiteness of the volume integral in $F_Q(s,T,\mu_B)$.
In combination with  the power $T$ dependence of  the reduced surface tension 
coefficient $\sigma(T)$
the same inequality provides   the existence and finiteness of 
all its partial  $T$ derivatives of $F_Q(s,T,\mu_B)$ regardless  to the sign of 
$\sigma(T)$.  Thus, using the Taylor expansion in powers of $(T - T_{cep})$ 
at any point of the interval
$T =  T_{cep}$ and $\mu_B < \mu_B^c(T_{cep})$, one can calculate 
$s_H(T , \mu_B)$ for the values of  $T > T_{cep}$ which are  inside the convergency  radius of  the Taylor expansion.  

The other situation is for $\mu_B \ge \mu_B^c(T_{cep})$ and $T > T_{cep}$,
namely
in this case above 
the deconfinement PT there must exist a weaker PT 
induced by the disappearance of the reduced  surface tension coefficient. 
To demonstrate this we have solve Eq. (\ref{s*vdw}) in the limit, when $T$ approaches 
the curve $T= T_{cep}$ from above, i.e. for  $T \rightarrow  T_{cep}+0$, and study the 
behavior of $T$ derivatives of the  solution of Eq. (\ref{s*vdw}) $s^*$ for fixed values of 
$\mu_B$.
For this purpose  we have to evaluate the  integrals ${\cal K}_\tau (\Delta,\gamma^2)$
introduced in Eq. (\ref{KQ}). 
Here  the notations 
$\Delta \equiv s^* - s_Q(T, \mu_B)$ and $\gamma^2 \equiv - \sigma (T) > 0$ are  introduced for convenience. 

To avoid the unpleasant behavior for $\tau \le 2$ it is convenient to transform  (\ref{KQ}) 
further on by integrating by parts: 
\begin{align}\label{KQ2}
{\cal K}_\tau (\Delta,\gamma^2) \, \equiv & ~ g_\tau(V_0)  - \frac{\Delta}{(\tau-1)} {\cal K}_{\tau -1} (\Delta,\gamma^2) + \nonumber \\
&\frac{\varkappa \, \gamma^2}{(\tau-1)} {\cal K}_{\tau-\varkappa} (\Delta,\gamma^2) \,,
\end{align}
where the regular function $g_\tau(V_0) $ is defined as
\begin{align}\label{gtau}
&  g_\tau(V_0) \equiv  \frac{1}{(\tau-1)\, V_0^{\tau-1}} \exp\left[ -\Delta V_0 + 
\gamma^2 V_0^{\varkappa}\right] \,. 
 \end{align}
For $\tau - a > 1$ one can change the variable of integration 
$v \rightarrow z / \Delta$ and  rewrite $ {\cal K}_{\tau-a} (\Delta,\gamma^2)$ as 
\begin{align}
\hspace*{-0.25cm}
 {\cal K}_{\tau- a} (\Delta,\gamma^2)  = \Delta^{\tau- a-1} \hspace*{-0.15cm} 
 \int\limits_{V_0 \Delta}^\infty  \hspace*{-0.15cm}
dz ~\frac{\exp\left[- z  + \frac{\gamma^2}{\Delta^\varkappa} z^{\varkappa} 
\right] }{z^{\tau- a}}   \equiv &
\nonumber \\
\Delta^{\tau-a-1} \, {\cal K}_{\tau-a} \left(1, \gamma^2\Delta^{-\varkappa} \right) \,.&
\label{KQ3}
 \end{align}
This result shows that  in the limit $\gamma \rightarrow 0$, when the rightmost 
singularity must  approach $s_Q(T,\mu_B)$ from above, i.e. $\Delta \rightarrow 0^+$, the function (\ref{KQ3}) behaves as  $ {\cal K}_{\tau- a} (\Delta,\gamma^2) \sim \Delta^{\tau-a-1} + O(\Delta^{\tau-a})$. This is so because for $\gamma \rightarrow 0$ 
the ratio $\gamma^2\Delta^{-\varkappa}$ cannot go to infinity, otherwise  the function 
${\cal K}_{\tau-1} \left(1, \gamma^2\Delta^{-\varkappa} \right) $,  which enters into the right hand side of  (\ref{KQ2}),   would diverge exponentially 
and this   makes impossible an existence of the solution of Eq. (\ref{s*vdw}) for 
$T = T_{cep}$. The analysis shows that for $\gamma \rightarrow 0$  there exist  two possibilities: either  $\nu \equiv \gamma^2\Delta^{-\varkappa} \rightarrow Const$ or 
$\nu \equiv \gamma^2\Delta^{-\varkappa} \rightarrow 0$.
The most straightforward  way to analyze these possibilities for  $\gamma \rightarrow 0$ is to assume the following behavior 
\begin{align}
&  \Delta  =  A\, \gamma^\alpha +  O(\gamma^{\alpha+1}) \,, 
\label{Dasgamma} \\
& \frac{ \partial \Delta}{\partial T}   =   \frac{ \partial \gamma}{\partial T} 
 \left[A\,\alpha\,  \gamma^{\alpha-1} +  O(\gamma^{\alpha})\right] 
 \sim \frac{(2\,k +1) A \,\alpha\,  \gamma^{\alpha}}{2\, (T - T_{cep}) },
\label{DTDasgamma} 
\end{align}
and find out the $\alpha$ value by equating (\ref{DTDasgamma}) with  the $T $ derivative (\ref{sHprime1}). 

Indeed, using  (\ref{sHprime1}), (\ref{G1}) and (\ref{KQ}), one can write
\begin{align}\label{Dsprime}
& \hspace*{-0.25cm}
\frac{ \partial \Delta}{\partial T}   = 
 \frac{ G_2  + 
 u\, {\cal K}_{\tau-\varkappa}(\Delta,\gamma^2) \,2 \,\gamma \gamma^\prime}{ 1 + u\, {\cal K}_{\tau-1}(\Delta,\gamma^2)} \approx 
 \frac{ \Delta^{2-\tau} G_2} {  u\, {\cal K}_{\tau-1}(1, \nu )}  
 + \nonumber \\
&  \frac{  
2\, \gamma \gamma^\prime
 \Delta^{1-\varkappa} \left[ \nu\, \varkappa \, {\cal K}_{\tau-2\varkappa}(1, \nu) - 
 {\cal K}_{\tau-1-\varkappa}(1, \nu) \right]}{ (\tau-1-\varkappa)\, {\cal K}_{\tau-1}  (1, \nu )  }\,,
 \end{align}
where the prime denotes the partial $T$ derivative. Note that  the function
$G_2 \equiv F^\prime+u^\prime {\cal K}_\tau(\Delta,\gamma^2)-s_Q^\prime $ can vanish for a few  values of $\mu_B$ only.
In the last step of deriving (\ref{Dsprime}) we used the  identities (\ref{KQ2}) and (\ref{KQ3})
and dropped the non-singular terms. As we discussed above, in the limit 
$\gamma \rightarrow 0$  the function $\nu$ either remains a constant or vanishes, then the term 
$ \nu\, \varkappa \, {\cal K}_{\tau-2\varkappa}(1, \nu) $ in (\ref{Dsprime})
is either of the same order 
as the  constant $ {\cal K}_{\tau-1-\varkappa}(1, \nu)$ or vanishes. 
Thus, to reveal  the behavior of  (\ref{Dsprime}) for $\gamma \rightarrow 0$  it is sufficient to find a leading term out of $\Delta^{2-\tau}$ and $\gamma \gamma^\prime  \Delta^{1-\varkappa} $ and compare it with the assumption (\ref{Dasgamma}).

The analysis shows that for  $\Delta^{2-\tau} \le  \gamma \gamma^\prime  \Delta^{1-\varkappa}$ the last term in the right hand side of (\ref{Dsprime}) is the leading one. 
Consequently, equating the powers of $\gamma$ of the leading terms in 
(\ref{DTDasgamma}) and (\ref{Dsprime}), one finds
\begin{align}\label{alpha1}
&  
%
\gamma^{\alpha -2} \sim \Delta^{1-\varkappa} \Rightarrow ~ \alpha \varkappa = 2 ~~{\rm for}~~ \tau \le 1 + \frac{\varkappa}{2 k + 1} \,, 
 \end{align}
where the last inequality follows from the fact that the term  
$\gamma \gamma^\prime  \Delta^{1-\varkappa}$  in (\ref{Dsprime}) is the dominant one.

Similarly, for  $\Delta^{2-\tau} \ge  \gamma \gamma^\prime  \Delta^{1-\varkappa}$ one 
obtains $\gamma^{\alpha -1} \gamma^\prime \sim \Delta^{2 -\tau}$ and, consequently,
\begin{align}\label{alpha2}
& 
\alpha  = \frac{2}{(\tau-1)(2k+1)} ~~{\rm for}~~ \tau \ge 1 + \frac{\varkappa}{2 k + 1} \,.
 \end{align}

Summarizing our results for $\gamma \rightarrow 0$ as
\begin{eqnarray}\label{DsdTtot}
\frac{ \partial \Delta}{\partial T}   \hspace*{-0.0cm}  \sim 
{\textstyle \frac{T_{cep}\gamma^{\alpha }}{T-T_{cep}} } =
\left\{
\begin{tabular}{ll}
\vspace{0.1cm}  ${\textstyle \left[ \frac{T-T_{cep}}{T_{cep}}\right]}^{\frac{2k+1}{\varkappa}-1} $\,, &  \hspace*{-0.2cm} $ \tau \le 1 + \frac{\varkappa}{2 k + 1}$\,, \\
& \\
${\textstyle \left[\frac{T-T_{cep}}{T_{cep}} \right]}^{\frac{2-\tau}{\tau-1}  } $\,, &  \hspace*{-0.2cm} $\tau \ge 1 + \frac{\varkappa}{2 k + 1} $\,,
\end{tabular}
\right.  \hspace*{-0.3cm}
\end{eqnarray}
we can also write the expression for the second derivative of $\Delta$  as 
\vspace*{-0.3cm}
\begin{eqnarray}\label{D2sdT2tot}
\frac{ \partial^2 \Delta}{\partial T^2}   \hspace*{-0.0cm}  \sim 
\left\{
\begin{tabular}{ll}
\vspace{0.1cm}  ${\textstyle \left[ \frac{T-T_{cep}}{T_{cep}} \right]}^{\frac{2k+1}{\varkappa}-2} $\,, &  \hspace*{-0.2cm} $ \tau \le 1 + \frac{\varkappa}{2 k + 1}$\,, \\
& \\
${\textstyle \left[ \frac{T-T_{cep}}{T_{cep}} \right]}^{\frac{3-2\tau}{\tau-1} } $\,, &  \hspace*{-0.2cm} $\tau \ge 1 + \frac{\varkappa}{2 k + 1} $\,.
\end{tabular}
\right.  \hspace*{-0.3cm}
\end{eqnarray}
The last result shows us that, depending on $\varkappa$ and $k$ values, 
the second derivatives of $s*$ and $s_Q(T,\mu_B)$ can differ from each other  for 
$ \frac{3}{2} < \tau < 2$ or can be equal for $ 1 < \tau \le \frac{3}{2}$.
In other words, we found  that at  the line $T = T_{cep}$ there exists  the 2$^{nd}$ order PT for $ \frac{3}{2} < \tau < 2$ and the higher order PT   for $ 1 < \tau \le  \frac{3}{2}$,
which separates the pure QGP phase from the region  of  a cross-over, i.e.  the mixed states of hadronic and QGP bags. Since it exists at the line of a zero surface tension, 
this PT  will be called the {\it surface induced PT.}

\vspace*{-0.0cm}

%
%
\begin{figure}[ht]
\includegraphics[width=8.4cm,height=6.3cm]{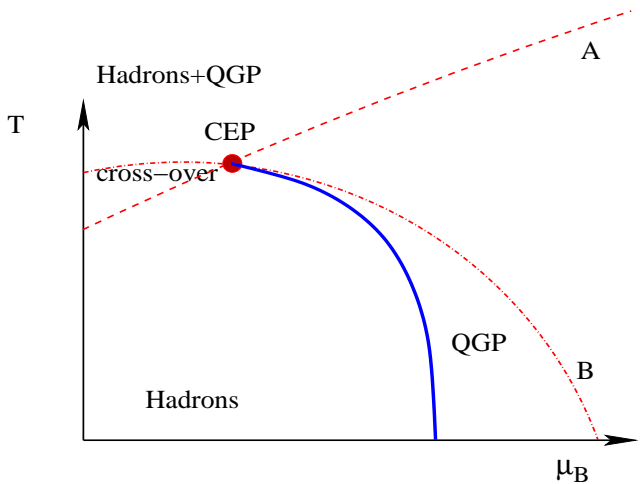}
\vspace*{-0.3cm}
\caption{
A schematic picture of the deconfinement phase transition diagram (full curve)
in the plane of  baryonic chemical potential $\mu_B$  and  $T$ for 
the 2$^{nd}$ order PT at the tricritical endpoint (CEP).
The model predicts an existence of the  surface induced PT of the 2$^{nd}$ or
higher order (depending on the model parameters). This PT starts at the CEP and goes 
to higher values of $T$ and/or $\mu_B$. Here it  is shown by the dashed curve CEP-A, if the phase diagram is endless, or
by the dashed-dot curve CEP-B, if the phase diagram ends at $T= 0$. 
Below (above) each of the dashed curves the reduced surface tension coefficient 
is positive (negative), 
}
  \label{fig5}
\end{figure}

\vspace*{-0.0cm}

Since the analysis performed in  the present section did not include any $\mu_B$ derivatives 
of $\Delta$, it remains valid for   the $\mu_B$ dependence of the 
reduced surface tension coefficient, i.e. for $T_{cep} (\mu_B)$.
Only it is  necessary to make a few comments on a possible location of  
the {\it surface tension null line}  $T_{cep} (\mu_B)$.  
In principle, such a null line can be located anywhere, if its location does not contradict 
to the sufficient conditions (\ref{SufCondI}) and (\ref{SufCondII}) of   the  1$^{st}$ deconfinement PT existence.  Thus, the surface tension null line must cross the 
deconfinement line in the $\mu_B-T$ plane at a single point which is  the tricritical endpoint
$(\mu_B^{cep}; T_{cep} (\mu_B^{cep})) $, whereas for  $\mu_B > \mu_B^{cep} $ the 
null line should have higher temperature for the same $\mu_B$   than the deconfinement one, i.e. 
$T_{cep} (\mu_B) > T_c  (\mu_B) $ (see Fig 5).  Clearly, there exist  two distinct cases
for the surface tension null line: either it is endless, or it ends at 
zero temperature. But recalling  that at low temperatures and high values of the baryonic chemical potential there may exist the Color-Flavor-Locked (CFL)  phase 
\cite{Krishna:98},
it is possible that the null line may also cross the boundary of the  CFL phase and, perhaps, it may create another special point at this intersection. 

One may wonder why this surface induced PT was not observed so far. 
The main reason is that the lattice QCD calculations at non-zero $\mu_B$ are very 
difficult, and  because of this   the identification  of the precise  location of the critical 
endpoint  is  highly nontrivial task \cite{fodorkatz,karsch,misha}. Therefore, 
the identification of the 2$^{nd}$ or higher order PT which might be located in 
the vicinity of  the deconfinement PT could be a real challenge. 
In addition, the surface induced PT  may lie so  close to the deconfinement line that 
it would be extremely difficult to observe it at the present lattices.


\vspace{-0.25cm}

\section{Conclusions and Perspectives}

\vspace{-0.25cm}

Here we suggested an analytically solvable  statistical model  which simultaneously  describes 
the 1$^{st}$ and 2$^{nd}$ order PTs with a cross-over. The approach is general and can be used 
for  more complicated parameterizations of  the hadronic mass-volume spectrum, if in the vicinity of
the deconfinement PT region the discrete and continuous parts  of this spectrum  can be expressed in the form of Eqs. (\ref{FHTmu}) and  (\ref{FQTmu}), respectively. Also the actual parameterization of 
the QGP pressure $p = T s_Q(T,\mu_B)$ was not used so far, which means that our result can be extended to more complicated functions, that can contain other phase transformations (chiral PT,
or the PT to color superconducting phase)   provided that the sufficient  conditions (\ref{SufCondI}) and (\ref{SufCondII})  for the deconfinement PT existence  are satisfied. 

In this model the desired properties of the deconfinement phase diagram are achieved by accounting for the temperature dependent surface tension of the quark-gluon bags. 
As we showed, it is crucial for the cross-over existence that at  $T= T_{cep}$ 
the reduced surface tension 
coefficient vanishes and remains negative for temperatures above  $T_{cep}$.
Then the deconfinement $\mu_B-T$  phase diagram has the 1$^{st}$ PT  at  
$\mu_B > \mu^c_B( T_{cep})$ for 
$ \frac{3}{2} < \tau <  2$ , which degenerates into the 2$^{nd}$ order PT 
(or higher order PT for  $ \frac{3}{2} \ge \tau >1 $)  at 
$\mu_B = \mu^c_B( T_{cep})$, and a cross-over for $0 \le \mu_B < \mu^c_B( T_{cep})$.
These two ingredients  drastically change the critical properties of the GBM 
\cite{Goren:81} and resolve the long standing problem of  a unified description of  the 
1$^{st}$ and 2$^{nd}$ order PTs and  a cross-over,
which, despite all claims,  was not resolved in Ref. \cite{Goren:05}. 
In addition, we found that at the null line of the surface tension there must exist 
the surface induced  PT of the 2$^{nd}$ or higher order, which separates 
the pure QGP from the mixed states of hadrons and QGP bags, that coexist above
the cross-over region (see Fig. 5).  Thus, the QGBST model predicts that the QCD critical endpoint is the tricritical endpoint. It would be interesting  to verify this prediction
with the help of the lattice QCD analysis. 
For this one will need to study the behavior of the bulk and surface contributions to 
the free energy of the  QGP bags  and/or  the string connecting 
the static quark-antiquark pair.

In contrast to  popular mean-field models the PT mechanism in  the present model is clear: it happens due to  the competition of the rightmost singularities of the isobaric partition function. Since the  GCP function of
the QGBST model does not depend on any  (baryonic or entropy or energy) density, but depends exclusively  on $T, \mu_B$ and $V$, its phase diagram  does not 
contain any back bending and/or spinodal instabilities  \cite{Randrup:04} which are typical for the mean-field (= classical) models.
The found exact analytical solution does not require a complicated and  artificial 
procedure of conjugating  the two parts of the equation of state  in the vicinity of the critical endpoint
like it is done by hands in Refs. \cite{Nonaka:05,Antoniou:05} because  all this is automatically included in the statistical description. 

Also in the QGBST model the pressure of the deconfined phase is generated by the infinite  bag,
whereas the discrete part of the mass-volume spectrum plays  an auxiliary role even above the cross-over region. Therefore,  there is no reason to believe that any quantitative changes of the properties of low lying hadronic states  generated by the surrounding  media (like the mass shift of the $\omega$ and $\rho$ mesons \cite{Shuryak:05}) would be the robust  signals of the deconfinement PT. 
On the other hand,  the QGP bags created in the experiments  have finite mass and volume and, hence,  the strong discontinuities  which are typical for the 1$^{st}$ order  PT should be smeared out 
which would  make them hardly distinguishable from the cross-over. Thus, to seriously discuss 
 the signals of the 1$^{st}$  order deconfinement PT and/or  the tricritical endpoint,  one needs to 
solve the  finite volume version of the QGBST model like  it was done for the SMM \cite{Bugaev:04a} and the GBM \cite{Bugaev:05c}.  This, however, is not sufficient because, in order to make any reliable prediction for experiments, the finite volume equation of state must be used  in   hydrodynamic equations which, unfortunately, are not suited for such a  purpose.  Thus, we are facing a necessity to return to the foundations of heavy ion phenomenology and to modify them according to the requirements of the experiments.  The present model can be considered as the next step in 
this direction.

Although the present model has a great advantage compared to other models because, in principle,  
it can be formulated on the basis of the experimental data on the degeneracies, masses and eigen volumes  of hadronic resonances in the spirit of Ref. \cite{Goren:82}, a lot of additional work is necessary to properly study the  issues addressed in \cite{Shuryak:06}. 
Thus, above the surface tension null line 
the hadrons can coexist with QGP  at high temperatures. Consequently, the nonrelativistic 
consideration of hard core repulsion in the present model should be modified to
its  relativistic treatment  for light hadrons like it is suggested  in 
\cite{Rvdw:1,Rvdw:2}.  This can lead to some new effects discussed recently in \cite{Rvdw:2}. 
Also,  the realistic equation of state requires the inclusion of the temperature and mass  dependent 
width of heavy resonances into a  continuous part of the mass-volume spectrum 
which  may essentially modify our understanding 
of the cross-over mechanism \cite{Blaschke:03}. 

Finally, a precise  temperature dependence of the  surface tension  coefficient of the bags 
should be investigated and its relation to the interquark string tension should be studied in detail. 
For this  it will be necessary  to modify  the Hills and Dales Model \cite{Bugaev:04b,Bugaev:05a}
in order to include the surface deformations with the base of  arbitrary size  whereas its present formulation is suited for  discrete clusters and, hence, for discrete bases of surface deformations.

\vspace{0.25cm} 

{\bf  Acknowledgments.}
I am thankful to A. Blokhin for important comments and   K. Rajagopal for pointing out   Ref. \cite{Jaffe}.
A warm hospitality of  the Frankfurt Institute for Advanced Studies, where 
an essential part  of this work was done, is appreciated. 
The financial support of the Alexander von Humboldt Foundation is 
acknowledged.



\end{document}